\documentclass[twocolumn]{aastex62}

\newcommand{\MSun}{\mbox{M$_\odot$}}
\newcommand{\RSun}{\mbox{R$_\odot$}}

\def\apgt{\ {\raise-.5ex\hbox{$\buildrel>\over\sim$}}\ }
\def\aplt{\ {\raise-.5ex\hbox{$\buildrel<\over\sim$}}\ }

\received{January 1, 2018}
\revised{January 7, 2018}
\accepted{\today}
\submitjournal{ApJ}

\shorttitle{Compact Binaries of BS Twins from Stellar Triples}
\shortauthors{Portegies Zwart \& Leigh}

\begin{document}

\title{A Triple Origin for Twin Blue Stragglers in Close Binaries}

\correspondingauthor{Nathan W. C. Leigh}
\email{nleigh@amnh.org}

\author{Simon Portegies Zwart}
\affiliation{Leiden Observatory \\
Leiden University \\
PO Box 9513, 2300 RA \\
Leiden, the Netherlands}

\author{Nathan W. C. Leigh}
\affil{American Museum of Natural History \\
Department of Astrophysics \\
79th Street at Central Park West \\
New York, NY 10024-5192, USA}
\affil{Stony Brook University \\
Department of Physics and Astronomy\\
Stony Brook, NY 11794-3800, USA}
\affil{Departamento de Astronom\'ia \\ 
Facultad de Ciencias F\'isicas y Matem\'aticas \\ 
Universidad de Concepci\'on \\ 
Concepci\'on, Chile}





\begin{abstract}

We propose a formation mechanism for twin blue stragglers (BSs) in
compact binaries that involves mass transfer from an evolved outer
tertiary companion on to the inner binary via a circumbinary disk.  We
apply this scenario to the observed double BS system Binary 7782 in
the old open cluster NGC 188, and show that its observed properties
are naturally reproduced within the context of the proposed model.
Based on this model, we predict the following properties for twin BSs:
(1) For the outer tertiary orbit, the initial orbital period should
lie between 220 days $\lesssim$ P$_{\rm out}$ $\lesssim$ 1100 days,
assuming initial masses for the inner binary components of $m_{\rm 1}
= 1.1$ M$_{\odot}$ and $m_{\rm 2} =$ 0.9 M$_{\odot}$ and an outer
tertiary mass of $m_{\rm 3} = 1.4$ M$_{\odot}$.  After Roche-lobe
overflow, the outer star turns into a white dwarf (WD) of mass 0.43 to
0.54\,\MSun. There is a correlation between the mass of this WD and
the outer orbital period: more massive WDs will be on wider orbits.
(3) The rotational axes of both BSs will be aligned with each other
and the orbital plane of the outer tertiary WD. (4) The BSs will have
roughly equal masses, independent of their initial masses (since the
lower mass star accretes the most).  The dominant accretor should,
therefore, be enriched more effectively by the accreted material.  As
a result, one of the BSs will appear to be more enriched by either He,
C and O or by s-process elements, depending on if the donor started to
overflow its Roche lobe on, respectively, the red giant or asymptotic
giant branch.  (5) Relative to old dense clusters with high-velocity
dispersions, twin BSs in close binaries formed from the proposed
mechanism should be more frequent in the Galactic field and younger
open clusters with ages $\lesssim$ 4-6 Gyr, since then the donor will
have a radiative envelope. (6) the orbit of the binary BS will have a
small semi-major axis (typically $\aplt 0.3$\,au) and be close to
circular ($e \aplt 0.2$).

\end{abstract}


\section{Introduction} \label{intro}

Blue straggler stars are brighter and bluer than the main-sequence
(MS) turn-off in a cluster colour-magnitude diagram
\citep[e.g.][]{1953AJ.....58...61S,1989AJ.....98..217L,2014ApJ...782...49S}.
Two primary channels for BS formation have been proposed: mass
transfer from an evolved donor on to a MS star in a binary star system
\citep[e.g.][]{1964MNRAS.128..147M,1997A&A...328..143P,2009Natur.457..288K,2011MNRAS.410.2370L,2011Natur.478..356G},
and direct stellar collisions involving MS stars likely mediated via
binaries
\citep[e.g.][]{1975AJ.....80..809H,1997A&A...328..130P,2007ApJ...661..210L,2013MNRAS.428..897L,2013MNRAS.429.1221H, 2019A&A...621L..10P}.
The first mechanism predicts BSs in binaries with WD companions,
whereas the second predicts MS companions in a wide and eccentric
binary.  Other possible, albeit related, formation mechanisms include
mergers of close MS-MS binaries \citep{2019A&A...621L..10P}, and
mergers of the inner binaries of hierarchical triple star systems
induced by Lidov-Kozai oscillations coupled with tidal damping
\citep[e.g.][]{2009ApJ...697.1048P}.

In spite of these specific predictions for the expected properties of
BSs formed from each of the above production mechanisms, many BSs
exist with observed properties that defy these simple scenarios.  For
example, in the old open cluster (OC) M67, there lurks a candidate triple
system that is posited to host two BSs
\citep{2001A&A...375..375V,2003AJ....125..810S}.  The observations
suggest that the outer tertiary is itself a BS, with a mass $\sim$ 1.7
M$_{\odot}$ and orbiting the inner binary with a period of $\sim
1188.5$ days \citep{2003AJ....125..810S}.  The inner binary has a
period of only $\sim 1.068$ days \citep{2001A&A...375..375V}, and
hosts a BS of mass $\sim 2.52$ M$_{\odot}$.  In order to reproduce the total system mass 
we require at least five stars \citep{2011MNRAS.410.2370L}.  This is strongly indicative of a
dynamical origin for the system, and a single direct interaction
involving a binary and a triple that resulted in two separate
collisions is the most probable explanation for its origin
(instead of back-to-back direct binary-binary interactions)
\citep{2004MNRAS.350..615G,2011MNRAS.410.2370L}.  

Even more curious, there exists in the old OC NGC 188 a
double BS binary, called Binary 7782.  The BS
population in NGC 188 has a bi-modal period-eccentricity distribution.
As discussed in \citet{2011MNRAS.410.2370L}, this could be hinting at
a triple origin for at least some subset of the total BS population.
As for Binary 7782, \citet{2009Natur.462.1032M} observed a compact and
mildly eccentric (i.e., $e \sim 0.1$) binary star system with an
orbital period of $\sim$ 10 days hosting two roughly equal-mass BSs.  During a given binary-binary interaction, the probability
that not one but two direct (MS-MS) collisions will occur is less than
$10^{-2}$
\citep{1989AJ.....98..217L,2011MNRAS.410.2370L,2012MNRAS.425.2369L}.
Plus, binaries with collision products typically have relatively long
orbital periods \cite{2011Sci...334.1380F}. Dynamically, it is
difficult to form a short-period binary composed of two collision
products during a collisional interaction in a star cluster
\citep{2011MNRAS.410.2370L,2011Sci...334.1380F}.  So, how did Binary 7782 form?

We propose a formation channel for Binary 7782, and compact double BS
binaries in general, which involves mass transfer from an outer
tertiary companion on to an inner binary composed of two MS stars.  In
section~\ref{sect:dyn}, we constrain the range of initial (i.e.,
pre-mass transfer) orbital parameters for a hypothetical outer
tertiary companion, using a combination of dynamical and stellar
evolution-based constraints.  In Section~\ref{sims} we present the
numerical simulations used to study the mass transfer process in this
triple system. We adopt orbital parameters that, according to our
expectations, are most promising for the progenitors of the twin BS
7782.  The calculations are performed using the Astrophysical
Multipurpose Software Environment \cite[\texttt{AMUSE} for short,
  see][]{PortegiesZwart2013456,AMUSE} with a combination of stellar
evolution, hydrodynamical and gravitational simulations.  With these
calculations we further constrain the possible range of initial
parameters that naturally lead to twin BSs with orbital parameters
similar to the 7782 system.  We summarize and discuss the implications
of our results for compact double BS binaries and, more generally,
mass transfer in stellar triples in Section~\ref{sect:discussion}.

\section{Constraints on the present-day orbital parameters for a hypothesized
         tertiary companion in the compact BS Binary 7782} \label{sect:dyn}

\begin{figure}[ht!]
\includegraphics[width=\columnwidth]{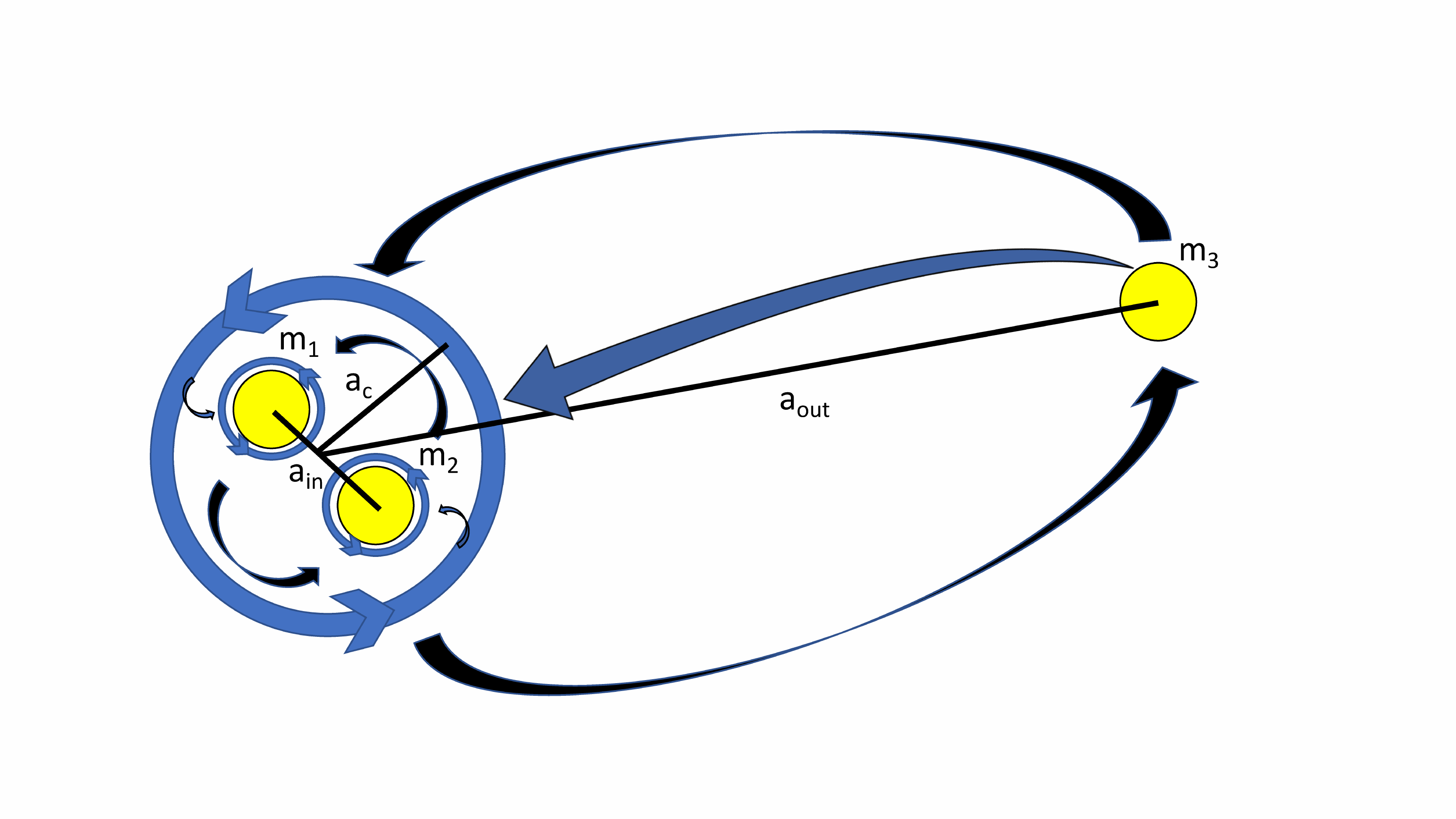}
\caption{Cartoon depiction of our proposed scenario for the formation
  of Binary 7782, specifically mass transfer from an evolved outer
  tertiary companion on to a compact inner binary via a circumbinary
  disk.  The outer tertiary component has mass m$_{\rm 3}$, whereas
  the inner binary components have masses m$_{\rm 1}$ and m$_{\rm 2}$.
  The inner and outer orbital separations are denoted by,
  respectively, a$_{\rm in}$ and a$_{\rm out}$.  The circularization
  radius of the accretion stream is denoted a$_{\rm c}$, as calculated
  via Equation~\ref{eqn:ac}, and marks the mean separation of the
  circumbinary disk.
\label{fig:fig1}}
\end{figure}

In our scenario, we start with a binary star with component masses
$m_{\rm 1}$ and $m_{\rm 2}$ that is orbited by a tertiary of mass
$m_3$. The inner and outer binary orbital semi-major axes are denoted
a$_{\rm in}$ and a$_{\rm out}$, respectively.  For clarity we 
assume both orbits, the inner as well as the outer, to have negligible
eccentricity and low inclination.  These assumptions are also
supported by the population of observed low-mass triples
\citep{2010yCat..73890925T,2018ApJ...854...44M}.  This initial configuration for our
assumed formation scenario for Binary 7782, described below, is
depicted in figure~\ref{fig:fig1}.

According to our scenario $m_3 > m_1 > m_2$ and the outer orbit is
sufficiently small that the tertiary star is filling its Roche lobe
and transfers mass to the inner binary before it leaves the asymptotic
giant branch. We constrain the inner orbit by requiring the triple
system to be dynamically stable, for which we adopt eq.\,1 in
\cite{1999ASIC..522..385M}.  While transferring mass, the accretion
stream gathers around the inner binary at the circularization radius
a$_{\rm c}$, and forms a circumbinary disk
\citep{2002apa..book.....F}.  Using conservation of angular momentum,
we equate the specific angular momentum of the accreted mass at the
inner Lagrangian point of the (outer) donor star to the final specific
angular momentum of the accretion stream at the circularization radius
about the inner binary, this results in
\begin{equation}
\label{eqn:specangmom1}
v_{\rm orb,3}(a_{\rm out} - R_{\rm L}) = v_{\rm orb,c}a_{\rm c},
\end{equation}
where R$_{\rm L}$ is the radius of the Roche lobe of the outer
tertiary companion, $a_{\rm c}$ is the semi-major axis of the orbit
about the inner binary corresponding to the circularization radius and
v$_{\rm orb,c}$ is the orbital velocity at $a_{\rm c}$.  The distance
from the centre of mass corresponding to the tertiary defined by the
Roche lobe is given by eq.\,2 in \cite{1983ApJ...268..368E}.
Combining eq.\,2 in \citet{1983ApJ...268..368E} (with mass ratio q
$=$ m$_{\rm 3}$/(m$_{\rm 1} +$m$_{\rm 2}$)) with
eq.~\ref{eqn:specangmom1}, we solve for the circularization
radius as a function of a$_{\rm out}$ and the assumed stellar masses:
\begin{equation}
\label{eqn:ac}
a_{\rm c} = a_{\rm out}(1 - R_{\rm L}).
\end{equation}
In order for a circumbinary disk to form around the inner binary, we
require that a$_{\rm in} <$ a$_{\rm c}$.

\begin{figure}[ht!]
\includegraphics[width=\columnwidth]{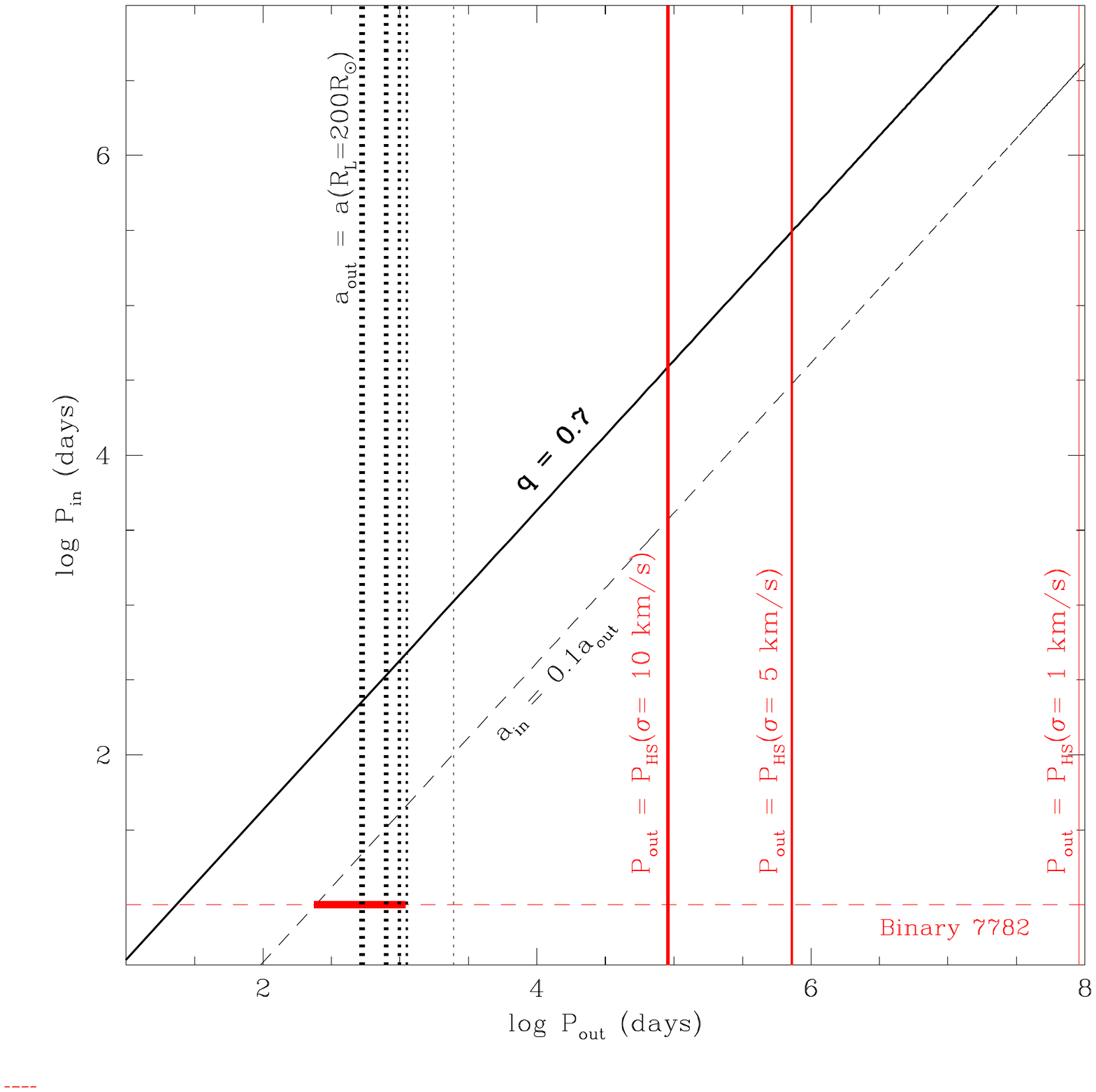}
\caption{Parameter space in the P$_{\rm out}$-P$_{\rm in}$-plane
  allowed for the hypothetical outer tertiary orbit of Binary 7782
  before Roche-lobe overflow.  The solid diagonal black line shows the
  period corresponding to the circularization radius a$_{\rm c}$ for
  the mass transfer stream coming from the outer star (i.e., at the
  onset of mass transfer).  
We assume initial component masses of m$_{\rm 1}$ = 1.1 M$_{\rm
  \odot}$ and m$_{\rm 2}$ = 0.9 M$_{\rm \odot}$ for the inner binary
components, and m$_{\rm 3}$ is computed for the outer tertiary
according to our assumed mass ratio (with our fiducial case
corresponding to q $=$ 0.7).  We assume completely conservative mass
transfer for this exercise, and a final mass for the outer tertiary of
0.6 M$_{\rm \odot}$ once it has become a WD.  The dashed diagonal
black line shows a rough criterion for dynamical stability in the
triple, approximately following \citet{1999ASIC..522..385M} (i.e.,
a$_{\rm in} <$ 0.1a$_{\rm out}$ is required for long-term dynamical
stability in equal-mass co-planar triples).  The vertical solid red
lines show the outer orbital periods corresponding to the hard-soft
boundary assuming central velocity dispersions of $\sigma =$ 1, 5 and
10 km s$^{-1}$.  The vertical dashed black lines show the maximum
outer orbital period P$_{\rm out}$ for which the outer tertiary
companion is Roche lobe-filling, assuming a stellar radius of R$_{\rm
  3} =$ 200 R$_{\odot}$ (which corresponds to the maximum stellar
radius reached on the AGB for the range of tertiary masses of interest
to us; see figure~\ref{fig:tertiarymass_vs_size}).  The horizontal
dashed red line shows the observed orbital period for Binary 7782,
using its observed orbital period and our assumed final inner
companion masses (i.e., m$_{\rm 1} =$ m$_{\rm 2} =$ 1.4 M$_{\odot}$).
Finally, the thick solid horizontal red line shows the parameter space
for P$_{\rm out}$ allowed after considering all of the aforementioned
criteria.
\label{fig:fig2}}
\end{figure}

Figure~\ref{fig:fig2} shows the parameter space in the P$_{\rm
  out}$-P$_{\rm in}$-plane for Binary 7782.  We assume initial
component masses of $m_{\rm 1} = 1.1$ M$_{\rm \odot}$ and $m_{\rm 2} =
0.9$ M$_{\rm \odot}$ for the inner binary components, and $m_{\rm 3} =
1.4 $M$_{\rm \odot}$ for the outer tertiary.  We compare the
circularization radius to the semi-major axis of the inner binary, for
which we require $a_{\rm c} > a_{\rm in}$, after folding in all
constraints from the requirements for dynamical stability (listed in
the caption of figure~\ref{fig:fig2}), and the assumption of an outer tertiary that is Roche
lobe-filling.  
Note that the range of
plotted orbital periods P$_{\rm in}$ corresponding to a contact state
for the inner binary lies outside the range of plotted values for
$P_{\rm in}$ (for components with radii of 1 R$_{\odot}$),
since it does not contribute to constraining the outer orbital
properties.  The thick horizontal solid red line shows the allowed
range of outer semi-major axes, after folding in all of 
the aforementioned criteria.  These constraints result in a rather narrow
range of initial conditions for the outer orbit, namely 2.2 $\times$
10$^{2}$ days $\le$ P$_{\rm out}$ $\le$ 1.1 $\times$ 10$^3$ days.

\begin{figure}[ht!]
  \includegraphics[width=\columnwidth]{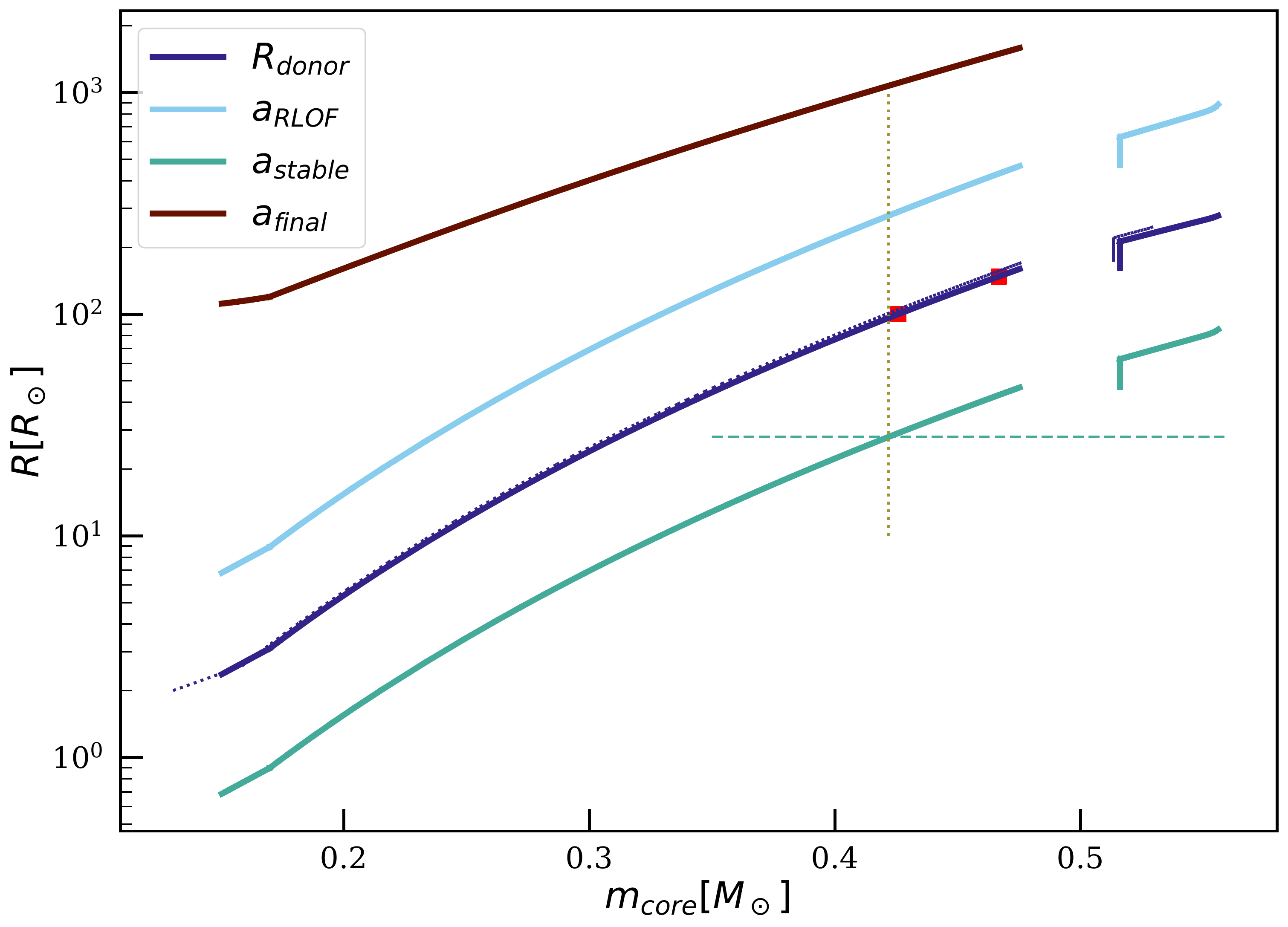}
  \caption{Giant radius as a function of the mass of the core of the
    Roche-lobe filling outer star (dark blue curve).  Here we adopt a
    donor mass of 1.4\,\MSun, but for an 1.2\,\MSun\, the donor the
    curve is quite similar (see dotted dark-blue curve).  The red
    squares in the curve show the parameters for which we performed
    more detailed gravitational-hydrodynamical simulations (see
    \S\,\ref{sims}).  The horizontal dashed line shows the orbital
    separation of the observed twin BS 7782.  The initial triple in
    which it possibly formed must at least have been dynamically
    stable. The minimal orbital separation for the inner binary for
    which the triple is stable is given by the lower green coloured
    curve.  Donors which are smaller than about 100\,\RSun\,
    (light-green curve indicated with a$_{\rm stable}$) result in a
    dynamically unstable triple. The minimal core mass associated with
    a stable triple is then indicated by the left-most vertical dotted
    line.  The orbital separation at which the donor star overfills
    its Roche lobe is indicated with the light-blue curve. The top
    curve (brown) shows an estimate of the final orbital separation of
    the outer star, and therefore of the final orbit of the WD around
    the inner twin BSs.  For core masses $\apgt 0.5$\,\MSun\, the
    final orbital separation, after mass transfer, is smaller than the
    initial orbit.  Here we adopted an initial inner binary mass of
    (1.0+0.9)\,\MSun\, and a final twin BS mass of (1.4+1.4)\,\MSun.
\label{fig:tertiarymass_vs_size}}
\end{figure}

Adopting a mass for the tertiary star of $m_3 = 1.4$\,\MSun, we can
constrain the initial parameters for the inner binary as well as the
orbit of the outer star after mass transfer. We first calculate the
stellar radius as a function of core mass. In
figure\,\ref{fig:tertiarymass_vs_size} we present this relation
calculated using the {\tt SeBa} stellar evolution code
\citep{1996A&A...309..179P} as the dark blue curve.  The interruption
in this curve, around a core mass of $m_{\rm core} \sim 0.5$\,\MSun\,
is a result of the evolution along the horizontal branch, where the
core of the star continues to grow but the radius actually shrinks.
Roche-lobe overflow in this phase is not expected to happen, because
it would already have happened in an earlier evolutionary state of the
donor star, when it was bigger.

Adopting masses for the inner binary $m_1=1.1$\,\MSun\, and
$m_2=0.9$\,\MSun\, we can calculate the outer orbital separation at
the onset of Roche-lobe overflow $a_{\rm out}$, and subsequently the
maximum orbital separation for the inner binary for which the orbit is
stable and a circumbinary disk can form. These two limits are
presented as the light blue and light green curves in
figure\,\ref{fig:tertiarymass_vs_size}.  The allotted region of
parameter space is then above the dashed horizontal line and to the
right of the vertical dotted line.

With the adopted parameters, we can also estimate the final orbital
period of the left-over core from the tertiary star after 
mass transfer.  The change in orbital separation due to
non-conservative mass transfer can be expressed in terms of the mass
of the outer star before and after mass transfer, i.e. $m_3$ and $m'_3$
respectively, the total mass in the inner binary before ($m_{\rm in}$)
and after accretion ($m'_{\rm in}$) and the amount of angular momentum
lost per unit mass $\eta \simeq 3$. Adopting the relation between the
orbital separation before mass transfer ($a$) and after mass transfer
($a'$) from \cite{1995A&A...296..691P}
\begin{equation}
  {a' \over a} = \left( {m_3 m_{\rm in} \over m'_3 m'_{\rm in}} \right)^{-2}
  \left( {m_3 + m_{\rm in} \over m'_3 + m'_{\rm in}} \right)^{2\eta + 1},
\end{equation}
we arrive at the top brown curve in
figure\,\ref{fig:tertiarymass_vs_size}. This curve provides a prediction for 
the current orbital separation of the WD around the twin
BS 7782. Tidal effects during mass tranfer
have probably circularized the orbit, although some slight eccentricity
due to turbulent motion in the outer layers of the donor star may have
induced a small $e \ll 0.1$ eccentricity.

Having limited parameter space for the formation of the twin BS 7782, we continue by performing a series of simulations to
investigate the accretion and changes to the inner orbits of triple
systems in this range of parameters.

\section{Numerical Simulations} \label{sims}

We perform simulations of a triple star system for which the outer
star overfills its Roche lobe while the inner binary remains
detached. The calculations start by evolving the three stars to the
same age, which is selected such that the outer-most star fills its
Roche lobe.  First order constraints for the initial conditions are
derived in the previous \S. In the following two sections we describe
how we set up these simulations and then discuss the results. The
calculations are performed using the Astrophysical Multipurpose
Software Environment using a combination of stellar evolution,
gravitational dynamics and hydrodynamics.

\subsection{Setting-up the simulations}

We adopt initial masses of $m_{\rm 1} = 1.1$ M$_{\rm \odot}$ and
$m_{\rm 2} = 0.9$ M$_{\rm \odot}$ for the inner binary components, and
between $m_{\rm 3} = 1.2$ and $m_{\rm 3} = 1.4$ M$_{\rm \odot}$ for
the tertiary star.  We evolve the tertiary star using the MESA
stellar-evolution code \cite{2011ApJS..192....3P} to a radius of about
100\,\RSun\, and 150\,\RSun, at which point we assume it to overfill
it's Roche lobe (see red square in
figure~\ref{fig:tertiarymass_vs_size}).  We perform calculations for an
inner orbital separation of $a_{\rm in} = 0.10$\,au, $a_{\rm in} =
0.15$\,au and $a_{\rm in} = 0.20$\,au.  In total we performed 12
calculations at a resolution of 40k SPH particles and 12 at 80k.

The stellar-evolution model, including the structure, temperature and
composition profiles are turned into a smoothed-particles
representation using the module {\tt StellarModelInSPH} in AMUSE (see
chapter 4 in \cite{AMUSE}).  We follow the same procedure as described
in \cite{2014MNRAS.438.1909D} for simulating the future of the triple
system $\chi$ Tau (HD 97131) in which the outer-most star overfills
its Roche lobe and transfers mass to an inner binary.  After
generating the hydrodynamical representation of the donor star we
replace the stellar core by a point mass to prevent the majority of
the resolution to be confined in the star's central regions.  In a
following step we relax the star using the hydrodynamics solver. This
relaxation process is realized in 100 steps during which we reduce the
velocity dispersion of individual SPH particles to a glasses structure 
(see, for example, \S\,3.3 on page 40 in White (1995)).  During this procedure, the gaseous
envelope of the star tends to expand by about 20\%.  To determine the
radius of the evolving star we calculate Lagrangian radii and use the
distance to the stellar center which contains 90\% of its mass. From
this 90\% mass-radius relation we obtain the stellar radius and match
it with the Roche-lobe of the outer orbit.

\begin{figure}[ht!]
  \includegraphics[width=\columnwidth]{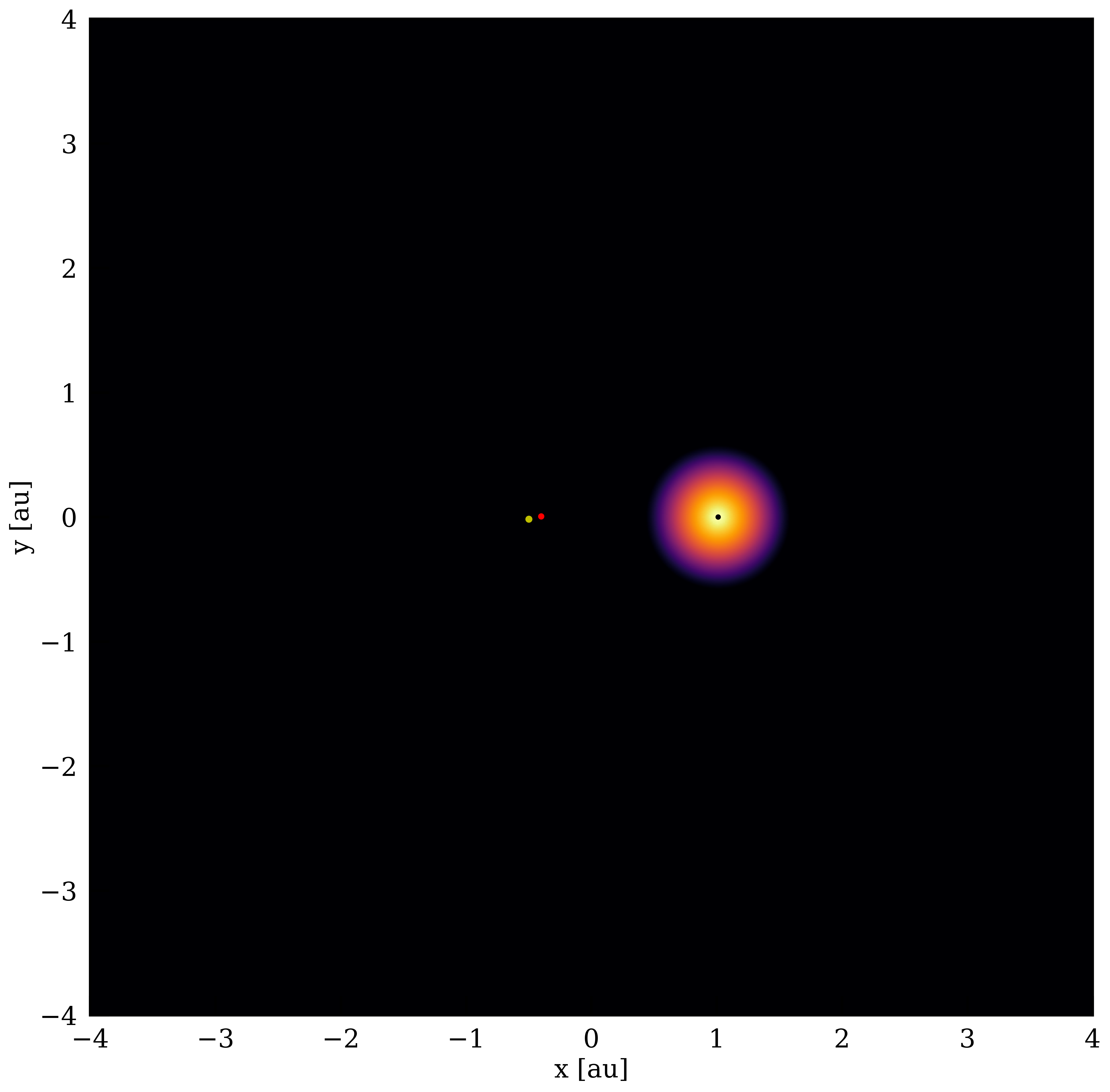}
\caption{Top view of the simulated triple system in which the
  100\,\RSun\, outer star of 1.4\,\MSun\, over-fills its
  Roche-lobe.  The star is represented by 80000\,SPH particles and a
  core particle of $\sim 0.4$\,\MSun (black bullet). The two companion
  stars are represented as black bullets (to the right).  The inner
  binary is represented by the yellow and red bullets for, respectively, the
  1.1\,\MSun\, primary and 0.9\,\MSun\, secondary stars in a circular
  orbit of 0.1\,au. The 1.4\,\MSun\, giant star is presented to the
  right in a circular orbit with semi-major axis $\sim 250$\,\RSun\, in the plane of the
  inner binary.
\label{fig:topview_at_t0}}
\end{figure}

With these parameters the orbital separation of the outer binary
becomes $\sim 250$\,\RSun for the 100\,\RSun donor star and about
430\,\RSun\, for the more evolved donor star.  We adopt the outer
orbit to be circular and in the plane of the inner binary.
In figure~\ref{fig:topview_at_t0} we present a top view of the initial
conditions for one of these calculations.

Roche-lobe overflow in triples is modelled using a coupled integrator
to follow the complex hydrodynamics of mass transfer from the
Roche-lobe filling outer star to the inner binary, while keeping track
of the gravitational dynamics of the stars.  The equations of motion
of the inner binary are solved using the symplectic direct N-body
integrator \texttt{Huayno} \citep{2012NewA...17..711P}. The
hydrodynamics are performed with the smoothed-particles hydrodynamics
code \texttt{Gadget2} \citep{2000ascl.soft03001S}, using an adiabatic
equation of state.  The two inner binary stars are treated as point
masses, but we allow them to accrete mass and angular momentum from
the gas liberated by the outer star.  This is realized using spherical
sink-particles that co-move with the mass points in the gravity
code. While the inner two stars accrete mass, they also accrete the
corresponding amount of angular momentum from the gas (see chapter 5
in \cite{AMUSE}).  The N-body integrator correctly accounts for this.
For the radius of the sink particles, we adopt $2 R_\odot$ for both
stars.

The N-body code, as well as the hydrodynamics solver, operate using
their own internal time-steps. The coupling between the two codes is
realized using the \texttt{Bridge} method in the AMUSE framework
\citep[see Sect.\.4.3.1 in][]{2013CoPhC.183..456P}.  This coupled
integrator is based on the splitting of the Hamiltonian, much in the
same way as is done with two different gravity solvers by
\cite{2007PASJ...59.1095F}. With the adopted scheme, the
hydrodynamical solver is affected by the gravitational potential of
its own particles, as well as the gravitational potential of the inner
binary. The hydrodynamics affects the orbits of the two inner stars
and the accretion onto the two stars affects the hydrodynamics. With
\texttt{Bridge} we realize a second order coupling between the
gravitational dynamics and the hydrodynamics.  The interval at which
the gravity and hydrodynamics interact via \texttt{Bridge} depends on
the parameters of the system we study, but typically we achieve
converged solutions when this time step is about 1/100 that of the
inner binary orbital period.

\section{Results of the hydrodynamical simulations} \label{results}

\begin{figure*}[ht!]
  \includegraphics[width=0.5\linewidth]{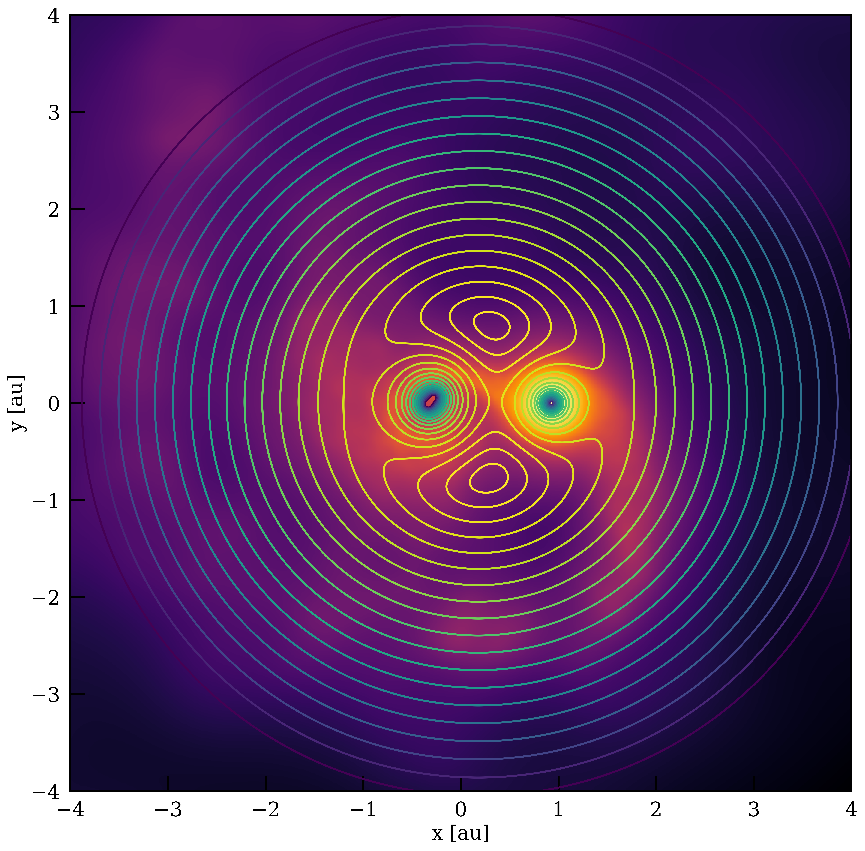}
~  \includegraphics[width=0.5\linewidth]{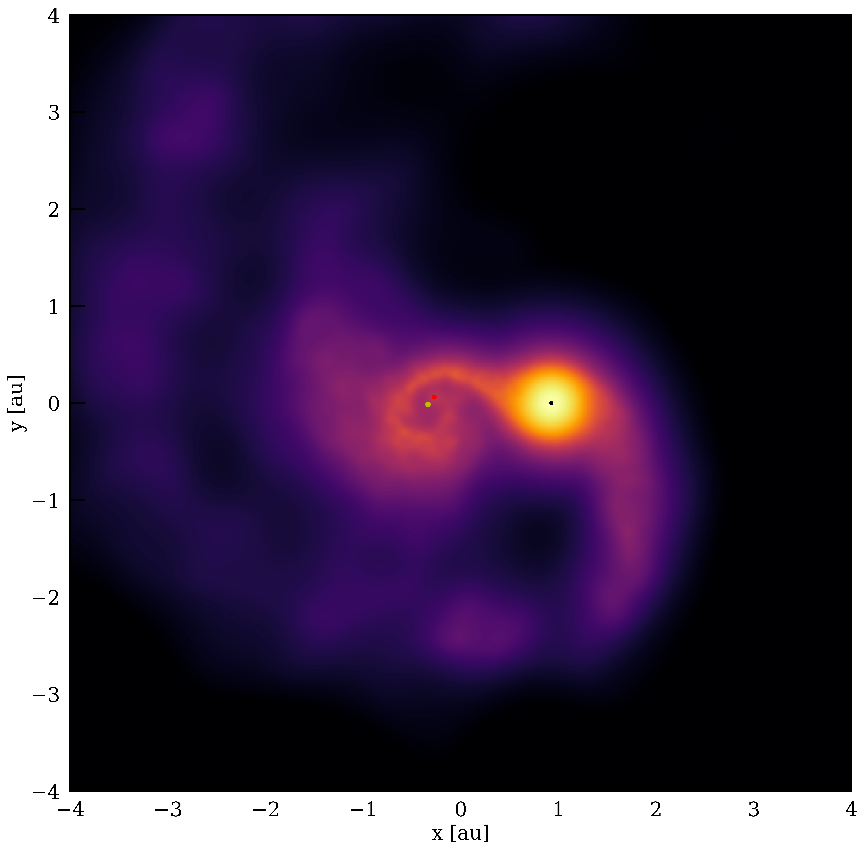}
\caption{Same as figure~\ref{fig:topview_at_t0} but at an age of $t
  \simeq 1091\,days$ after the start of the simulation. Left panel
  shows the equipotential surfaces of the triple overplotted with the
  gas distribution, the right panel shows just the gas and the stars
  as bullets.
\label{fig:topview_at_t1000day}}
\end{figure*}

\begin{figure*}[ht!]
  \includegraphics[width=0.51\linewidth]{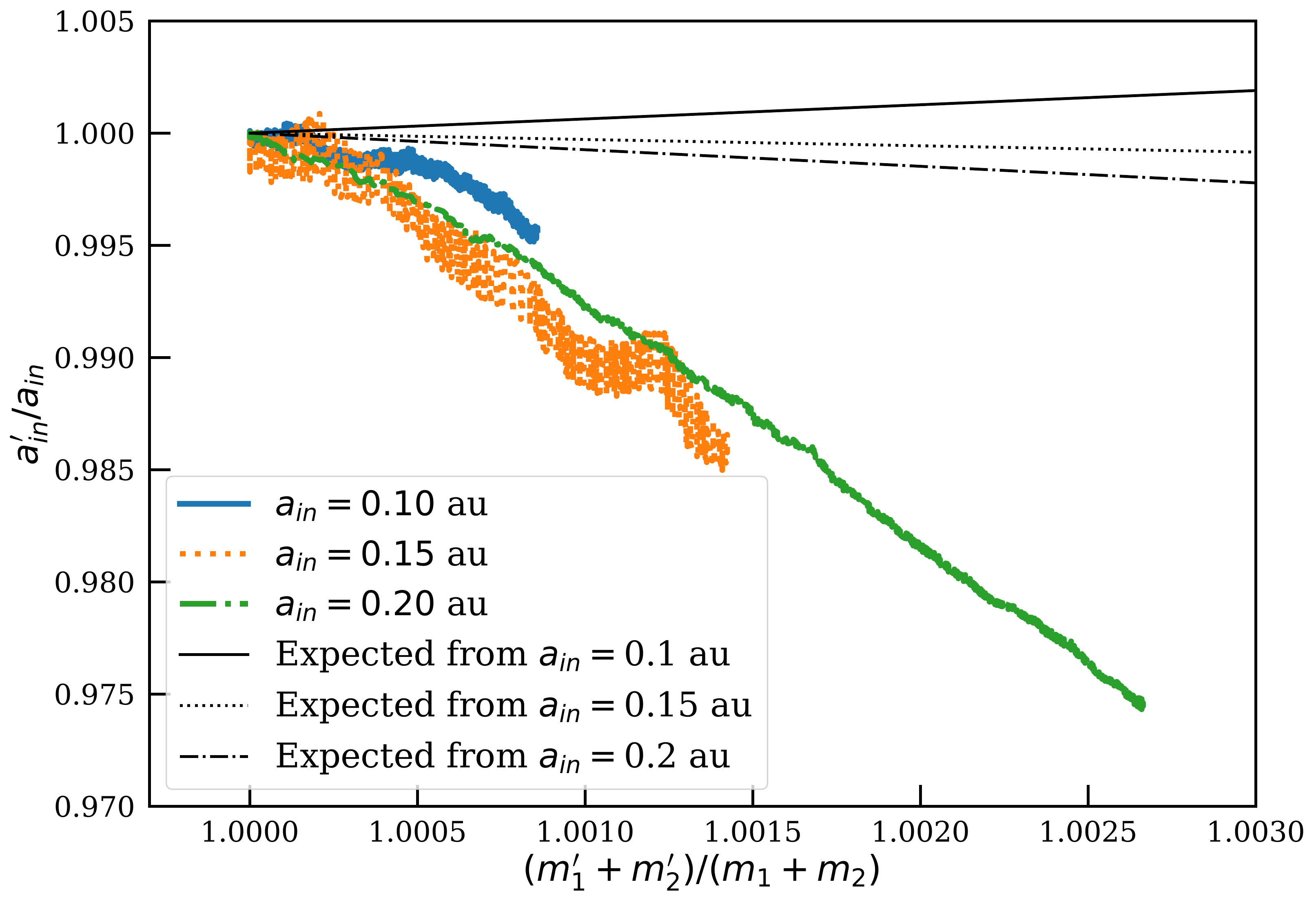}
~  \includegraphics[width=0.49\linewidth]{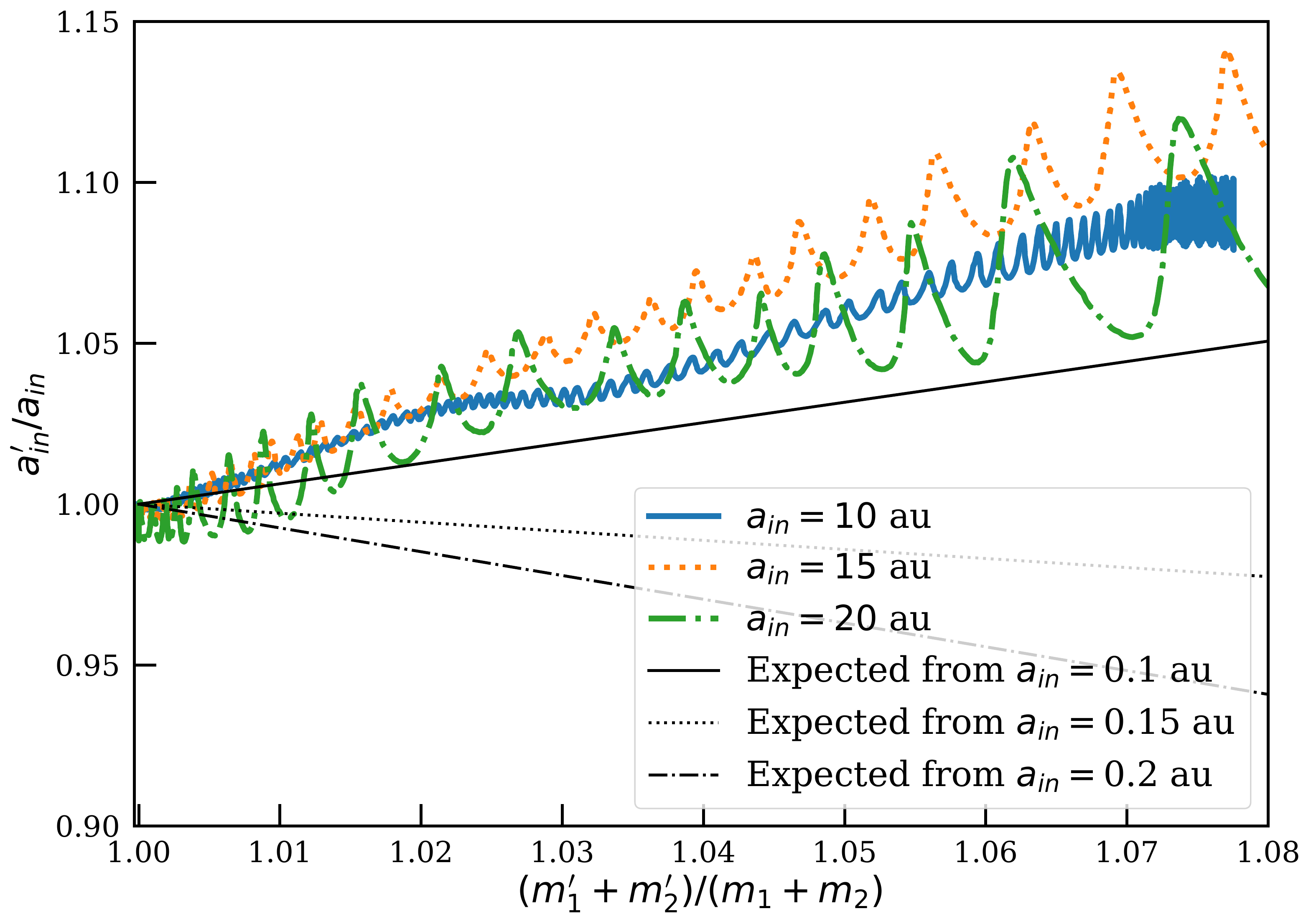}
  \caption{Evolution of the orbital separation as a function of the
    total mass of the inner binary for six calculations with somewhat
    different initial conditions (see the legend). The left panel
    shows the result for a 1.2\,\MSun\, donor star and the right panel
    for a 1.4\,\MSun\, donor. The initial binary shown by the blue
    curve of the right-hand panel is presented in
    fig.\,\ref{fig:topview_at_t0}, and in
    fig.\,\ref{fig:topview_at_t1000day} we present the final
    conditions of this system.  The black curves give the expected
    evolution of the orbital separation of the inner binary assuming
    that the binary evolved towards to observed orbital separation of
    0.13\,au at a total binary mass of 2.8\,\MSun.
\label{fig:mass_vs_semimajor_axis}}
\end{figure*}

To test the hypotheses that (1) the secondary in the inner binary
accretes more effectively than the primary star and to measure the
change to the inner orbit due to the Roche-lobe overflow of the outer
star, we perform a series of calculations in which we take the self
gravity and the hydrodynamical effects of the triple into account.
The results of these simulations are presented in
figure~\ref{fig:topview_at_t1000day} and
figure~\ref{fig:mass_vs_semimajor_axis}.  The first figure
(figure~\ref{fig:topview_at_t1000day}) shows the top view of the same
initial realization for which we presented the initial conditions in
figure~\ref{fig:topview_at_t0} but now at an age of 1091 days after
the onset of mass transfer. We add, to the left panel,
the equipotential surfaces in the orbital plane.

It is apparent that the mass transfer in the adopted triples leads to
a rather untidy evolution, since much of the donor mass is lost
through the second Lagrangian point to the right side of the donor
star in figure~\ref{fig:topview_at_t1000day}. A considerable amount of
mass is also lost through the third Lagrangian point (to the left of
the inner binary), although it is hard to actually quantize the amount
of material los, becuase an appreciable fraction is expected to rain
back onto the triple system.  One remaining question is how much mass
is eventually ejected altogether from the triple system and is
therefore not accreted to any of the two inner stars. This value is
hard to estimate from the simulations, but an accretion efficiency of
$\apgt 0.6$ is necessary to make the scenario feasible. Over the time
scale for which we performed the calculations, this efficiency is
reached, but it is not clear how the system responds at later stages.

The evolution of the inner orbit presented for several simulations in
figure~\ref{fig:mass_vs_semimajor_axis} is complicated.  This is
caused by the complex transport of mass, energy and angular momentum
through the accretion stream and throughout the system.  It is
therefore hard to quantify distinct trends in the evolution of the
triple system. In simulations of the response of an inner binary on
accretion from a circumbinary disk, \cite{2018arXiv181208175M}
conclude that the complexity of angular momentum transport between the
outer star and the accretion stream onto the individual inner stars,
is complicated and without clear trends. For most of our calculations
we agree with this statement, but in
figure~\,\ref{fig:mass_vs_semimajor_axis} we nevertheless present the
results of 6 of our calculations, three for a 1.2\,\MSun\, donor star
and three for a 1.4\,\MSun\, donor. The various coloured curves give
the resulting evolution of the inner orbit as a function of the total
mass in the inner binary. As the inner two stars accrete, the orbit
shrinks for a 1.2\,\MSun\, donor. These systems are expected to result
in a contact binary, that eventually may merge to form a single BS
with a mass more than twice the turn off. The required evolution in
order to explain the observed twin BS 7782 is indicated by the three
black curves; the simulated path clearly deviates from these. We,
therefore, argue that a 1.2\,\MSun\, donor has difficulty explaining
the observed orbital separation of $\sim 0.13$\,au in BSS 7782.

In the right-hand panel in figure~\ref{fig:mass_vs_semimajor_axis} we
present the evolution of the orbit for the 1.4\,\MSun\, donor for
several initial orbits of the inner binary. A more massive donor
appears to be more effective in producing a twin BS with parameters
consistent with the observed system 7782. There is more mass available
in the envelope of the donor star, and the orbital evolution of the
inner binary matches better with the anticipated evolution.  A more
massive donor may therefore have a lower accretion efficiency while
still accomodating the observed constraints.  The longer thermal time
scale of the stellar envelope of the higher-mass donor at the same
stellar radius eventually leads to a higher mass-transfer rate, and
therefore to a lower accretion efficiency. However, the larger mass
budget in the envelope appears to compensate.

The orbit of the inner binary expands in these cases as a result of
accretion onto the inner two stars. All three cases for the
1.4\,\MSun\, donor presented in
figure~\ref{fig:mass_vs_semimajor_axis} the inner orbit expands at
about the same rate. Consequently, the inner binaries that start with
$a = 0.15$\,au and $a=0.20$\,au eventually become dynamically
unstable.  The binary with an initial separation of $0.10$\,au expands
to reach a separation of about 0.126--0.145\,au for final masses for
the inner two stars of 1.4\,\MSun, which is consistent with the
observed twin BS 7782. In our simulations the eccentricity of the
inner binary grows to about $e \simeq 0.0028$.

With the accretion of mass, both stars in the inner binary also
accrete angular momentum.  By the end of the simulation the spins of
the two BSs are aligned along the orbital angular momentum axis with
an angle of $90.0^\circ$ for the primary star and $93.4^\circ$ for the
secondary star with respect to the argument of pericenter of the inner
orbit.  By the end of the simulations the spin of the primary is about
50.5 rotations per day, and 41.5 rotations per day for the secondary
star.

\section{Discussion} \label{sect:discussion}

In this paper, we consider the formation of twin BSs in tight binaries.
These systems may form through mass transfer from an outer Roche-lobe
filling tertiary star. Once this star ascends the giant branch, part
of its envelope is transferred to the inner binary, and accreted by
the two inner stars which are still on the MS.

As illustrated via SPH simulations, the mass transfer stream forms a
circumbinary disk, from which the inner binary stars accrete, driving
the inner binary toward a mass ratio close to unity.  Our simulations
indicate that the inner binary orbital separation can decrease or
expand depending on the details of the transfer of mass and angular
momentum.  More work is certainly needed in order to fully understand 
mass transfer in triples.

We summarize the results of these simulations as follows: for a
1.2\,\MSun\, tertiary donor mass, we expect the inner two stars to
eventually merge and form a single BS. This reduces the system to a
binary with a primary BS and an outer WD in a relatively wide
orbit. Such a BS will distinguish itself from other BSs by potentially
being more than twice the turn-off mass in a star cluster.  An example
could be the $2.9\pm0.2$\,\MSun\, BS S1237 in the Galactic cluster M67
\citep{2016ApJ...832L..13L}. It is the primary of a $\sim 698$\,day
binary with an eccentric orbit of $\sim 0.10$.

With an original outer star of mass $\sim 1.4$\,\MSun, the inner orbit
tends to expand. This eventually leads to a dynamically unstable
system resulting either in a collision or in the ejection of
(probably) the lowest mass star. This evolution could result in a
single ejected BS, with the other BS left in a relatively close and
eccentric orbit with a WD (the left-over core of the tertiary star).
These ``imposter'' BS-WD binaries would in principle mimic what is
expected theoretically for BSs formed from mass transfer in binary
stars. If such a dynamical instability engages relatively late in the
mass-tranfer phase, the white dwarf (maybe with a little left-over
envelope) is expected to be ejected. This would lead to a relatively
wide twin blue-straggler binary and a single low-mass white dwarf.

When we adopt an inner orbit of $0.10$\,au\, the expansion eventually
matches the observed orbital separation (i.e., $0.13$\,au) of the
observed twin BS 7782 and the observed masses of the two stars of
about 1.4\,\MSun.

In order to study the T-tauri binaries V4046 Sgr and DQ Tau,
\cite{2011MNRAS.413.2679D} perform a series of 2D hydrodynamical
simulations of circumbinary disks.  These authors studied the two
observed T-tauri systems V4046 Sgr and DQ~Tau, to which we compare our
results here.  For V4046 Sgr, for which the two stars have comparable
masses as in our calculation for a circular orbit with a period of
only 2.4 days, they find that the inner binary accretes at a rate of
$\sim 0.028$\,\MSun/Myr.  For DQ~Tau, which is composed of lower-mass
stars ($m_1 = m_2 \simeq 0.55$\,\MSun) in an eccentric ($e\simeq
0.556$) orbit of $\sim 15.8$\,days, they find an accretion rate onto
the inner binary of $\sim 0.027$\,\MSun/Myr.  These values are in the
same range as in our calculations, which results in an accretion rate
for the inner binary of 0.027--0.058\,\MSun/Myr (i.e., the average
measured over a period of about 3000 days in our simulations).
Interestingly, however, \cite{2011MNRAS.413.2679D} find that the
primary star in V4046 Sgr accretes at an 8\% higher rate than the
secondary star, whereas in our case the secondary star accretes at a
higher rate than the primary star by about 1\% to 12\%.  Higher
accretion rates in the secondary star are realized for eccentric and
retrogade inner orbits. We performed an extra series of calculations
to further study this, but they all lead to the merger of the inner
binary.


\section{Summary} \label{sect:conclusions}

In this paper, we propose a formation scenario for twin equal-mass
blue stragglers in tight binaries, as observed for Binary 7782 in the
old OC NGC 188.  The proposed scenario involves mass transfer from an
evolved outer tertiary companion, part of this mass is accreted by the
inner binary via a circumbinary disk the rest escapes through the
second and third Lagrangian points in the potential of the triple
system.  Our scenario makes several predictions for the observed
properties of a hypothetical outer triple companion, now a WD.  These
are:

\begin{enumerate}

\item For the predicted outer tertiary orbit, the initial orbital
  period should lie between 220 days $\lesssim$ P$_{\rm out}$
  $\lesssim$ 1100 days, assuming initial masses for the inner binary
  components of $m_{\rm 1} = 1.1$ M$_{\odot}$ and $m_{\rm 2} = 0.9$
  M$_{\odot}$ and an initial outer tertiary mass of $m_{\rm 3} = 1.4
  $M$_{\odot}$.

\item Larger final WD masses, and hence larger core masses for the
  donor at the time of mass transfer should correspond to larger final
  outer orbital periods for the tertiary.  This is because the Roche
  radius is larger for larger outer orbital periods, such that the
  donor must evolve to larger radii, and hence core masses, before the
  onset of mass transfer. We expect the orbital separation to range
  from $\apgt 6.4$\,yr for a $\sim 0.42$\,\MSun\, white dwarf to
  $\apgt 11.2$\,yr for a $\sim 0.48$\,\MSun\, white dwarf.

\item For the inner binary, the rotational axes of both the BSs should
  be aligned with each other and the orbital plane of the outer
  tertiary WD.  This is because accretion onto the BS progenitors
  proceeds via an accretion disk, that forms at the circularization
  radius and that has an orbital plane aligned with that of the outer
  tertiary.

\item The BSs in the inner binary should have roughly equal masses,
  independent of their initial masses.  This is because it is the
  lowest mass object that typically accretes the fastest, since its
  orbital velocity and distance relative to the circumbinary disk is
  typically the lowest
  \citep[e.g.][]{2000MNRAS.314...33B,2012ApJ...749..118S,2017MNRAS.466.1170M}.
  The mass ratio of the inner binary, therefore grows to unity.  As a
  consequence, the initially lower mass MS star should accrete the
  most, and therefore be more enriched by accreted material.  This
  could be observable in the surface layers of a radiative star.  If
  the donor is an RGB star, the accretors will be enriched in mostly
  carbon, oxygen and helium, but if the donor is an AGB star the
  enrichment will be mostly in s-process elements.
    
\item We expect twin BSs in compact binaries formed from the mechanism
  proposed here to be more frequent in younger clusters with ages
  $\lesssim$ 4-6 Gyr.  This is because clusters with a MS turn-off
  mass $\lesssim$ 1.2 M$_{\odot}$ have convective envelopes
  \citep[e.g.][]{1991ApJS...76...55I,2009pfer.book.....M}, and a
  radiative envelope for the donor in a mass transferring binary
  ensures stable accretion on to the accretor.  Note that part of the
  mass liberated from the triple system through the second and third
  Lagrangian points may eventually be accreted back onto the system.
  This could have interesting consequences for the enrichment of the
  low-mass white dwarf.


\end{enumerate}

We emphasize, in closing, that the choice for the initial mass of the
outer tertiary may be rather critical.  Mass transfer in our proposed
scenario proceeds from the most massive tertiary to a binary of lower
total mass. This may result in an unstable phase of mass transfer, in
particular if the donor has a convective envelope
\citep[e.g.][]{2009pfer.book.....M}. A radiative envelope of the donor
ensures that the mass transfer will be maximally conservative, such
that the accretion stream will be maximally stable, accreting at a
stable and roughly constant rate \citep[e.g.][]{1991ApJS...76...55I}.
This stability regime may also be of interest for explaining very
massive twins, of $\apgt 20$\,\MSun\, which could be promising sources
for gravitational wave detectors once both twins evolve to a binary
black hole \citep{2016MNRAS.460.3545D}.

\acknowledgments

N.W.C.L. acknowledges support from a Kalbfleisch Fellowship at the
American Museum of Natural History.  SPZ would like to thank Norm
Murray and CITA for their hospitality during my long-term visit.  This
work was supported by the Netherlands Research School for Astronomy
(NOVA). 
In this work we use the matplotlib
\citep{2007CSE.....9...90H}, numpy
\citep{Oliphant2006ANumPy}, AMUSE
\citep{portegies_zwart_simon_2018_1443252}, SeBa
\citep{2012ascl.soft01003P}, Huayno \citep{2012NewA...17..711P}, MESA
\citep{2010ascl.soft10083P}, and GadGet2 \citep{2000ascl.soft03001S}
packages. The calculations ware performed using the LGM-II (NWO grant
\# 621.016.701) and the Dutch National Supercomputer at SURFSara
(grant \# 15520).

\bibliographystyle{apj}
\bibliography{BSS7782}


\end{document}